\documentclass{optica-article}

\journal{opticajournal} 

\articletype{Research Article}

\usepackage{lineno,csquotes}
\usepackage{braket}

\begin{document}

\title{Multi-NARP Laser Driving Scheme for Multiplexed Quantum Networks}

\author{Ali Binai motlagh,\authormark{1,$\dag$} Grant Wilbur,\authormark{2,$\dag$}  Nour Allam,\authormark{2} Giannis Tolis,\authormark{2} Lilly Daw,\authormark{2} S. O’Neal,\authormark{3,$\ddagger$} Dennis G. Deppe,\authormark{3,$\mathsection$} and Kimberley C. Hall\authormark{2,*}}

\address{\authormark{1} Department of Applied Physics and Applied Mathematics, Columbia University, 500 W 120th St, New York, NY, 10027, US\\
\authormark{2} Department of Physics and Atmospheric Science, Dalhousie University, Halifax, Nova Scotia B3H 4R2, Canada\\
\authormark{3}The College of Optics and Photonics, University of Central Florida, Orlando, Florida 32816-2700, USA}

\authormark{$\dag$}The authors contributed equally to this work.\\
\authormark{$\ddagger$}Present address: IMEC, Kissimmee, FL 34744, USA\\
\authormark{$\mathsection$}Present address: SdPhotonics, Richardson, TX 75081, USA\\
\email{\authormark{*}Kimberley.Hall@dal.ca} 


\begin{abstract*} 

We extend the recently developed NARP scheme for laser-triggered single-photon sources to the simultaneous excitation of multiple emitters with varying transition energies, laying the groundwork for wavelength-division multiplexing in quantum optical networks. Our Multi-NARP scheme does not rely on polarization filtering and thus enables the near-unity extraction efficiency of single photons from each quantum dot.  Our approach also offers the advantages of robustness to variations in the laser pulse parameters and immunity to excitation-induced dephasing tied to electron-phonon coupling.  We show that simultaneous triggering of at least 10 emitters is possible, enabling the development of high-bandwidth quantum networks.
\end{abstract*}

\section{Introduction}
Single-photon sources (SPSs) are needed for many applications in quantum technology, including secure communication \cite{Kolodynski:2020} and the development of quantum networks \cite{Lu:2021}.   Among candidate SPSs, semiconductor quantum dots (QDs) offer on-demand operation and the potential for integration into chip-based quantum photonic circuits\cite{Rodt:2021,Zadeh:2016}. The performance of QD SPSs has continuously improved over the past 15 years \cite{Senellart:2017}, with a demonstrated photon indistinguishability of 99.5\% \cite{Somaschi:2016} and single-photon purity of 8$\times$10$^{-5}$ \cite{Schweickert:2018}.  QD SPSs with a high degree of photon number coherence have also been demonstrated \cite{Bozzio:2022,Karli:2024}, which are useful for quantum commmunication protocols such as twin field QKD. The quality of QD SPSs in the O- and C-bands, which are essential for fiber-based long-distance quantum communication applications \cite{Lu:2021,Wells:2023}, has also improved significantly over the past decade \cite{Wells:2023,Haffouz:2018,Vajner:2024,Kim:2025,Nawrath:2023,Anderson:2021}.  In recent years, QD SPSs have been used for quantum key distribution \cite{Bozzio:2022,Chaiwongkhot:2020,Zhang:2025}, quantum relays involving solid state nodes \cite{Deiteil:2017,Anderson:2020}, chip-based quantum computing\cite{Maring:2024} and quantum memories \cite{Thoma:2024}.  


The advent of new laser driving schemes for triggering QD-based SPSs has contributed to recent improvements in the quality of single-photon emission \cite{Glassl:2013,Reiter:2012,Huber:2020,Uppu:2020,Santori:2001,Mathew:2014,Wei:2014,He:2019,Koong:2021,Bracht:2022,Wilbur:2022}.  Among these schemes, those that implement shaping of the amplitude and/or phase of the excitation pulse are particularly attractive because they enable multiple performance metrics to be optimized simultaneously \cite{Mathew:2014,Wei:2014,He:2019,Koong:2021,Bracht:2022,Wilbur:2022}. Adiabatic rapid passage (ARP)-based approaches \cite{Reiter:2012,Mathew:2014,Wei:2014,Wilbur:2022}, which utilize a quadratic pulse phase, are especially promising for field-deployment of laser-triggered SPSs because they are robust to the laser pulse parameters (pulse area, frequency).  ARP also offers immunity of the quantum-state inversion process to phonon-based decoherence \cite{Mathew:2014,Kaldewey:2017,Ramachandran:2020} and allows for the direct preparation of the ground-state exciton in the QD, enabling the emission of highly indistinguishable single photons.  The recently proposed Notch-filtered Adiabatic Rapid Passage (NARP) scheme \cite{Wilbur:2022} is a modified version of ARP that combines these advantages with efficient extraction of the single photon emission.  Under direct ground-state exciton pumping, polarization-based filtering approaches are commonly used to separate the single photons from the scattered laser light \cite{Kuhlmann:2013}, but such approaches sacrifice half of the emitted photons,  impeding the realization of a deterministic photon source.  NARP circumvents this by enabling direct ground state pumping through adiabatic inversion in the presence of a spectral notch resonant with the exciton transition.  In conjunction with an efficient band pass filter on resonance with the transition, this approach reduces loss levels to a few percent or less \cite{Wilbur:2022}.  These promising laser driving schemes illustrate the utility of pulse shaping approaches for optimizing the performance of laser-triggered SPSs. 



Here we extend the NARP scheme to the parallel driving of multiple, spectrally distinct SPSs using a single laser pulse, supporting the development of high-bandwidth quantum networks through wavelength division multiplexing of single photon sources.  We refer to this approach as multi-NARP.   Multiplexing approaches using space \cite{Beraza:2025,Lio:2018,Koong:2020}, wavelength \cite{Eriksson:2019}, and orbital angular momentum states \cite{Zahidy:2022} are currently being explored to increase the bandwidth of quantum communication systems and to enable the development of new network strategies \cite{Park:2022}.    Our multi-NARP scheme exploits the robustness of ARP to the detuning of the laser pulse from the QD transition energy \cite{Ramachandran:2021} in conjunction with notched laser driving for efficient isolation of the emission from spectrally distinct optical channels.  We show that parallel quantum state initialization of at least 10 quantum emitters is feasible, extending our recent demonstration of ARP on 30 QDs \cite{Ramachandran:2024} to multi-NARP for triggering bright emitters. We demonstrate parallel quantum state initialization of two QDs experimentally using multi-NARP.  Our findings will support the development of high-performance quantum networks.

\section{Multi-NARP Laser Triggering Scheme}
\begin{figure}[ht!]
\centering
\includegraphics[width=12cm]{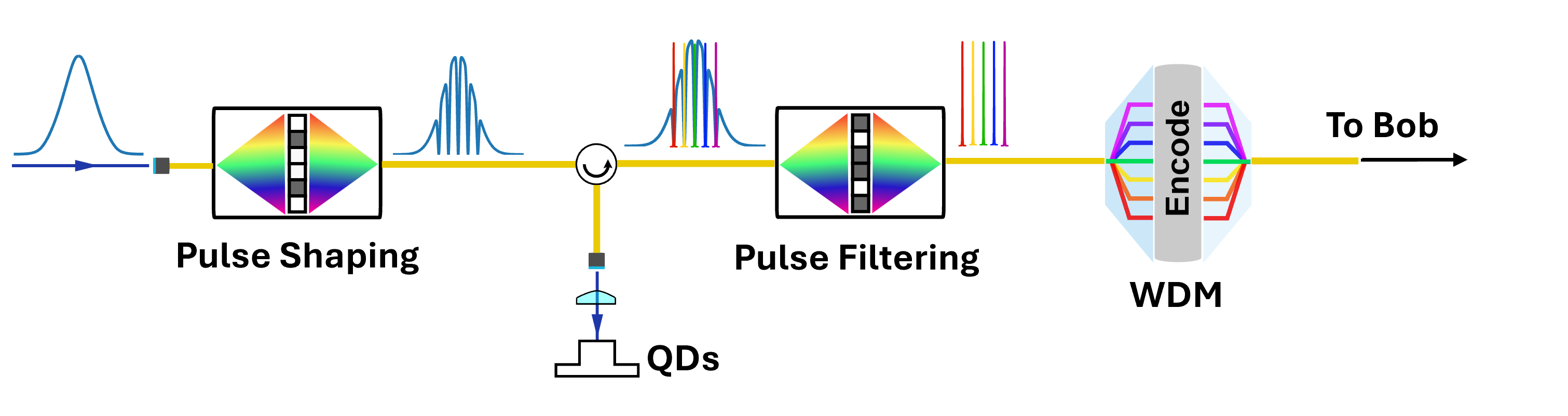}
\caption{Multi-NARP scheme, illustrating the simultaneous inversion of 5 quantum dots.  A transform-limited laser pulse is passed through a pulse shaping system, which imposes spectral chirp and notches centered on the transition energy of each quantum emitter.  This shaped laser pulse excites an ensemble of QDs, triggering photon emission into the 5 spectral channels in parallel.  A second pulse shaping system is used to filter out the scattered light from the exciting laser pulse, while transmitting single photons. Standard 4f pulse shapers or a compact chirped fiber Bragg grating in conjunction with notch and band pass filters could be used for the Pulse Shaping and Pulse Filtering stages. A wavelength division multiplexer (WDM) system is used to separate the channels for signal encoding and to recombine them into a single spatial mode before sending to Bob. Numerical calculations indicate that the channel capacity can be extended by at least an order of magnitude with this scheme.}
\label{multiNARP}
\end{figure}

Our multi-NARP approach utilizes pulse-shape engineering to realize a robust laser triggering scheme for parallel inversion of multiple spectrally distinct emitters, enabling high-bandwidth quantum communication through wavelength division multiplexing and large photon extraction efficiency. Multi-NARP is illustrated schematically in Fig.~\ref{multiNARP} for the case of five emitters.  Quantum dots with emission wavelengths that span the bandwidth of the laser pulse are selected from a dense ensemble \cite{Ramachandran:2024}, each of which serves as an on-demand source of single photons. The laser used to trigger the SPSs is sent through a pulse-shaping system that introduces chirp and spectral notches centered on the emission wavelength for each QD.  These QDs emit into a single spatial mode, which is then spectrally filtered to separate the emitted single photons from the scattered laser light.  The different frequencies are then demultiplexed and quantum information is encoded in specific modes of the electromagnetic field (e.g., polarization or time-bin). The frequency modes are finally combined into a single spatial mode before being sent to their destination (\textit{e.g.} other nodes of a quantum network). The implementation of WDM schemes in quantum networks would facilitate high-bit quantum communication and the integration of classical and quantum communication hardware \cite{Beraza:2025,Lio:2018,Eriksson:2019,Zahidy:2022}.  As we show below, high-fidelity quantum state inversion may be achieved for all QDs using the multi-NARP scheme.  Our approach is insensitive to fluctuations of the laser source (pulse area, frequency) and excitation-induced dephasing of the driven QD exciton tied to electron-phonon coupling.  For optimal performance, the multi-NARP trigger protocol would be implemented in conjunction with a low-loss spectral filtering scheme, and the QDs would be embedded in a photonic cavity or waveguide to improve the quality of the spatial mode in which the photons are emitted, enabling a large collection efficiency using a high NA objective \cite{Koong:2020}.

The pulse shape required for multiNARP may be implemented by passing a transform-limited pulse through a frequency domain pulse shaping system, which imposes a mask M($\omega$) = A($\omega$) $\exp{[i\Phi(\omega)]}$. The phase mask $\Phi(\omega)=\phi^{''}(\omega-\omega_0)^2/2$ introduces chirp, leading to a sweep of the instantaneous frequency versus time.  A($\omega$) = 1 corresponds to traditional ARP.  In this case, the time-dependent electric field is given by E($t$) = 1/2 E$_p$($t$)$\exp{[-i(\omega_L t + \alpha t^2)]}$ and the frequency sweep rate $\alpha$ is related to the spectral chirp $\phi^{''}$ by $\alpha = 2\phi^{''}/[\tau_0^4/((2ln(2))^2 + (2\phi^{''})^2]$, where $\tau_0$~=~120 fs is the transform-limited pulse duration for our system. For the excitation of N QDs using the multi-NARP scheme, the amplitude mask of ARP is replaced by A($\omega$) = $\prod_{i=1}^N [1-\exp{[-(\omega-\omega_i)^2/\delta^2]}]$ where $\omega_i$ and $\delta$ denote the center position and width of each notch in the spectrum, respectively. This extends NARP, which retains the robustness of ARP and eliminates the trade-off between SPS brightness and indistiguishability, to N emitters. The pulse shaping and filtering stages could be implemented using a standard 4f pulse shaping system or high-efficiency notch and band pass filters in conjunction with a chirped fiber Bragg grating \cite{Remesh:2023}.

For quantum state inversion via ARP-based schemes including multi-NARP, the system evolves within one of the two dressed states ($\ket{\psi_{+}}$ and $\ket{\psi_{-}}$) during the laser pulse, which are time-dependent admixtures of the bare QD states ($\ket{0}$ and $\ket{1}$).  Provided that the pulse area and chirp are large enough to reach the adiabatic regime \cite{Shore:book}, the system remains in a single-dressed state for all times during the laser pulse. Inversion of the exciton occupation occurs because the dressed state reverses character (i.e. evolves from $\ket{0}$ to $\ket{1}$) with the system occupying the upper (lower) dressed state for laser driving with negatively (positively) chirped pulses.  Since the condition for adiabatic evolution may be satisfied in the presence of fluctuations of the laser pulse area and detuning relative to the QD optical transition, the inversion process using ARP is highly robust and thus suitable for field-deployed applications.


\section{Parallel Quantum State Inversion using multi-NARP}

\begin{figure}[ht!]
\centering
\includegraphics[width=10.5cm]{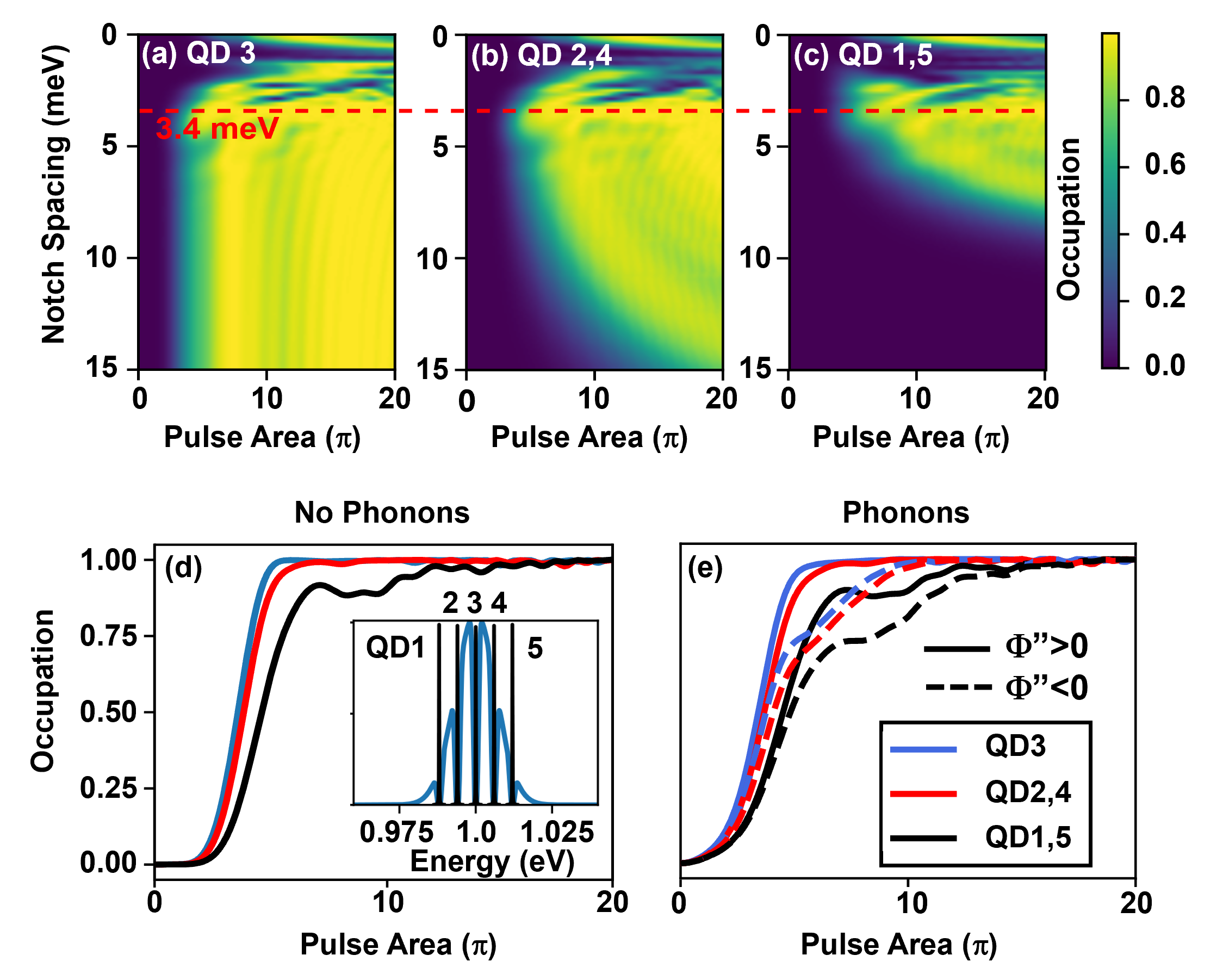}
\caption{Numerical simulation results for the simultaneous inversion of 5 QDs using multi-NARP. (a)-(c) Final occupation of the exciton in each QD versus pulse area and the spectral separation between the QD resonances. For these simulations, $\delta$~=~1~meV, $\tau_0$~=~120~fs, $\phi^{''}$~=~0.5~ps$^2$, and electron-phonon coupling was neglected. (d) Same results as (a)-(c) for a QD spacing of 3.4~meV.  Inset: Laser spectrum for a 3.4~meV hole spacing showing the labeling scheme used for the 5 QDs. (e) Calculated results with electron-phonon coupling included. Solid (dashed) curves show the occupation for a positively (negatively) chirped multi-NARP excitation pulse.}
\label{5hole}
\end{figure}

The quantum state dynamics for a system of spectrally distinct QDs excited using the multi-NARP scheme was calculated by integrating the optical Bloch equations, including electron-phonon coupling \cite{Mathew:2014,Ramsay:2011}.  The results of these simulations for the simultaneous driving of 5 QDs are shown in Fig.~\ref{5hole}.  For these calculations, $\tau_0$~=~120~fs, $\delta$~=~1~meV and $\phi^{''}$~=~0.5~ps$^2$, reflecting values commonly used in experiments \cite{Wilbur:2022}.  The QD transition energies are chosen to be symmetrically located with respect to the peak of the laser spectrum, and the QD dipole moments are set equal for all QDs for simplicity, but we obtained similar results for variations in these parameters (see Fig.~S2 in Supplement 1).   Fig.~\ref{5hole}(a)-(c) shows the final occupation of the exciton in each QD versus pulse area and the spacing between the QD resonances (\textit{i.e.} the notch spacing). Since the final occupation only depends on the magnitude of the detuning, QD 1 exhibits the same behavior as QD 5. The same is true for QDs 2 and 4. The driving laser spectrum and the labeling scheme used for the QDs are shown in the inset of Fig.~\ref{5hole}(d). High-fidelity quantum state inversion is achieved for all QDs provided the notches fall within the laser pulse bandwidth.  The exciton occupation versus the pulse area is shown in Fig.~\ref{5hole}(d) for the optimum notch separation of 3.4~meV. Robust inversion is achieved beyond the threshold pulse area for adiabatic state evolution, consistent with the results obtained previously for laser driving with ARP \cite{Mathew:2014,Wei:2014,Ramachandran:2021,Ramachandran:2024}.  In agreement with the findings for single NARP \cite{Wilbur:2022}, robust inversion occurs for multi-NARP despite the additional temporal structure tied to the spectral notches. This enables the advantages of ARP to be retained in conjunction with spectral isolation of photons emitted by all QDs from the scattered laser light for efficient photon extraction. The choice of notch width and laser pulse bandwidth together determine the number of QDs that may be simultaneously driven using multi-NARP.  Our simulations were extended to the inversion of 10 QDs in Fig.~S3 of Supplement 1, illustrating the versatility of the multi-NARP scheme.

Calculated results showing the role of excitation-induced dephasing tied to electron-longitudinal acoustic (LA) phonon coupling \cite{Mathew:2014,Ramsay:2011}, which impedes quantum state inversion for many proposed laser triggering schemes \cite{Glassl:2013,Reiter:2012,He:2019,Koong:2021,Bracht:2022,Vanucci:2023}, are shown in Fig.~\ref{5hole}(e). For chirped laser driving within the adiabatic regime at low temperatures, transitions caused by LA phonons can occur for negatively chirped pulses through phonon emission because in this case the system evolution occurs within the higher-energy dressed state.  Such transitions are suppressed for positively chirped pulses, for which the system remains in the lower-energy dressed state, due to the negligible phonon absorption probability caused by low phonon occupancy \cite{Mathew:2014,Kuhn:2012,Debnath:2012}. As a result, calculations of the dependence of the quantum state dynamics on the sign of the pulse chirp provide a clear picture of the role of phonons and the transition to the adiabatic regime.  For pulse areas larger than 5$\pi$, the dependence of quantum state inversion on pulse area is identical for the case of positively chirped pulses with phonons included (Fig.~\ref{5hole}(e), solid curves) and for electron-phonon coupling turned off (Fig.~\ref{5hole}(d)), indicating complete suppression of phonon-mediated dephasing during excitation. This ability to suppress excitation-induced dephasing through the use of a positively chirped multi-NARP pulse is a considerable advantage over competing laser trigger schemes. The separation between the dashed and solid curves in Fig.~\ref{5hole}(e) vanishes above 15$\pi$, indicating that the decoupling regime observed previously for traditional ARP \cite{Kaldewey:2017,Ramachandran:2020} is also accessible for multi-NARP.

\section{Experimental demonstration on Two Semiconductor Quantum Dots}

\begin{figure}[ht!]
\centering
\includegraphics[width=\textwidth]{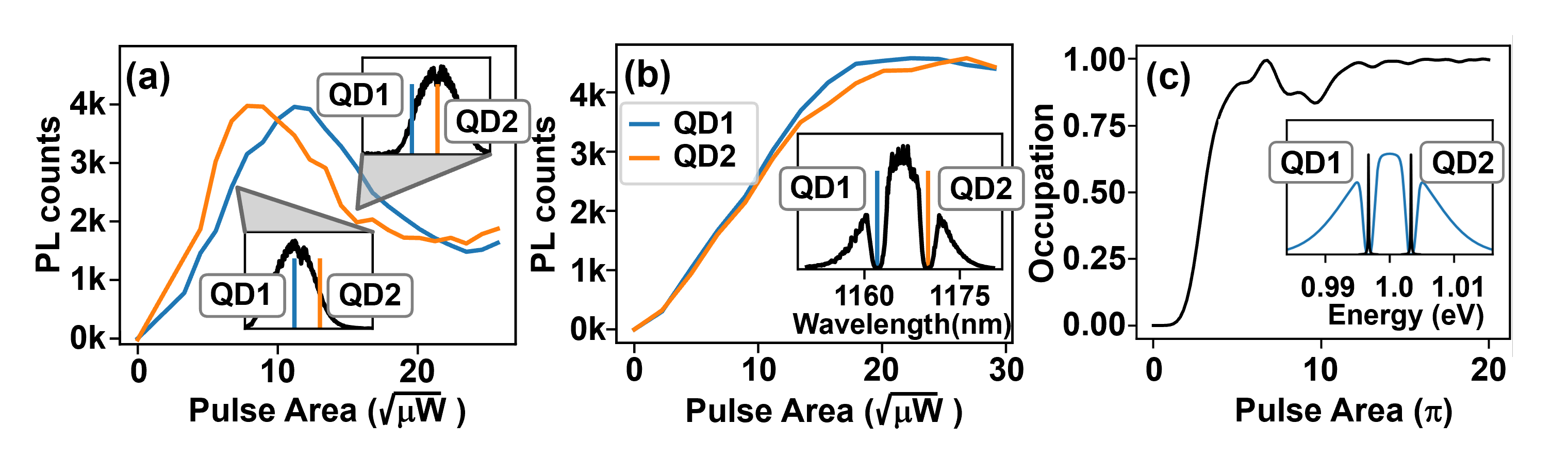}
\caption{Experimental demonstration of quantum state inversion using multi-NARP for two QDs.  (a) Measured Rabi rotation for resonant driving of the ES in each QD using transform-limited excitation pulses.  The laser tuning condition for each data set are shown as insets. (b) Experimental demonstration of simultaneous driving of QD~1 and QD~2 using multi-NARP.  Inset: Laser spectrum. (c) Calculated exciton occupation versus pulse area for multi-NARP with $N$~=~2, $\delta$~=~1~meV, $\tau_0$~=~120~fs, $\phi^{''}$~=~0.3~ps$^2$ and a notch separation of 7~meV to coincide with the experimental QD separation. In the simulations, equal dipole moments were used for QD 1 and QD 2 with symmetric transition energies relative to the peak of the laser spectrum.}
\label{Experiment}
\end{figure}

The results of experiments using the multi-NARP scheme on a pair of telecom-compatible InGaAs QDs are shown in Fig.~\ref{Experiment}.  The sample is a planar In(Ga)As/GaAs structure containing a single layer of QDs with an areal density of 1$\times$10$^{10}$~ cm$^{-2}$ \cite{Gamouras:2013}.  Multi-NARP is intended to drive the ground state (GS) transition, but quantum state inversion via multi-NARP was demonstrated on the first excited state (ES) transition (see Fig. S1) to simplify the experiments due to the low collection efficiency for the planar QD sample used.  In this configuration, the GS emission provides a proxy for the occupation of the ES at the end of the laser pulse \cite{Htoon:2002}.  

Fig.~\ref{Experiment}(a) shows the results of laser excitation using transform-limited pulses tuned into resonance with the ES transition in each QD, revealing a damped Rabi rotation.  For the demonstration of multi-NARP, the laser was tuned between the QD resonances, as shown in the inset of Fig.~\ref{Experiment}(b), and a quadratic phase mask with $\phi^{''}$~=~0.3~ps$^2$ was applied.  For multi-NARP excitation, the exciton inversion in both QDs first increases for low pulse area then transitions into a flat plateau, reflecting robust inversion.  The results of theoretical simulations of two-notch multi-NARP are shown in Fig.~\ref{Experiment}(d), also indicating a transition to the adiabatic regime.  More prominent fluctuations versus pulse area in the simulations, as seen previously with single NARP \cite{Wilbur:2022}, may reflect the differing shapes of the notches in theory and experiment.   

\begin{figure}[htbp]
\centering\includegraphics[width=8.7cm]{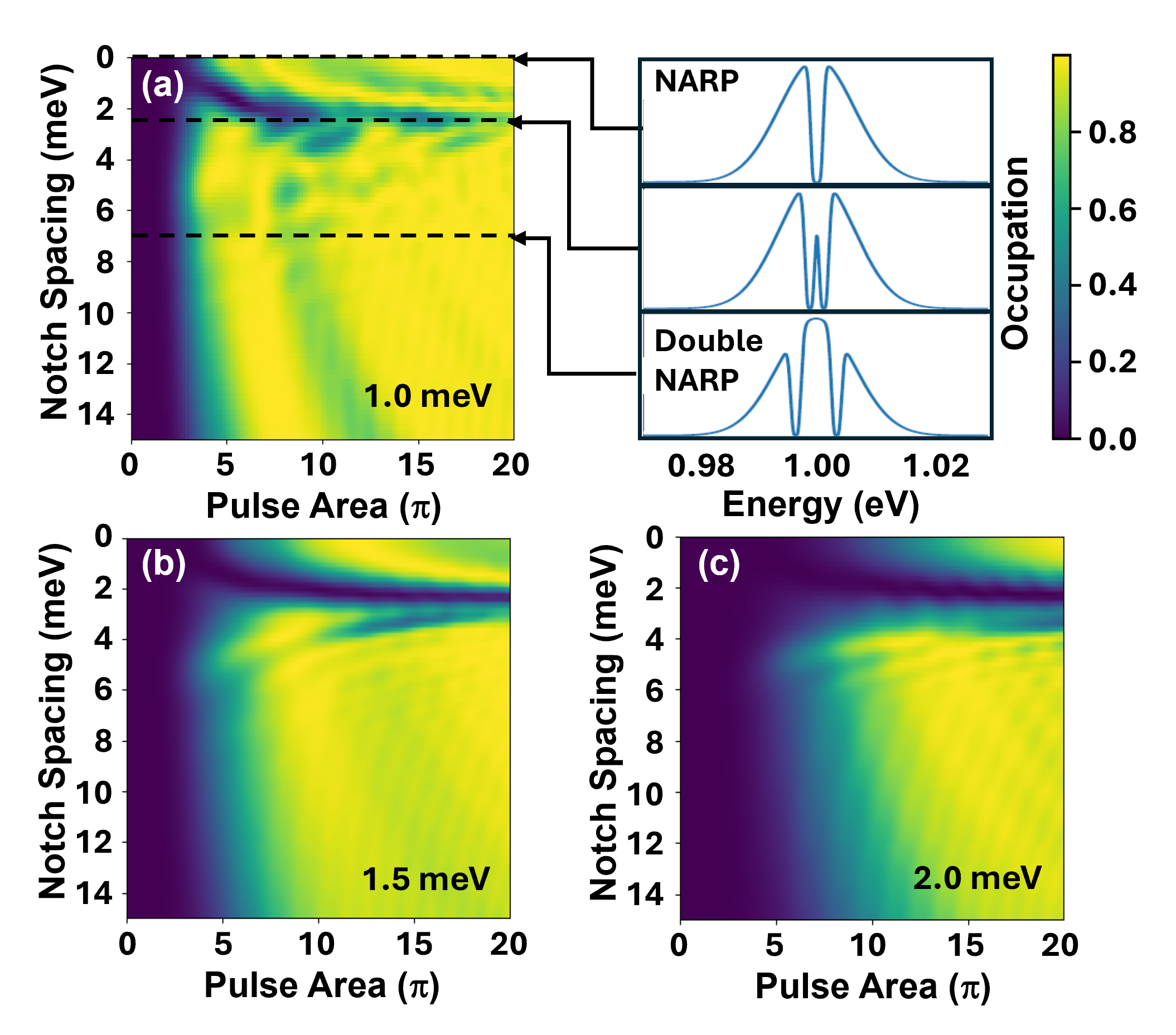}
\caption{Calculated exciton inversion versus pulse area and notch spacing for parallel driving of two QDs with multi-NARP, using a notch width of 1.0~meV [(a)], 1.5~meV [(b)], and 2.0~meV [(c)].  The inset to (a) shows the laser spectrum as the driving scheme transitions from parallel single NARP (top) to multi-NARP (bottom).}
\label{holewidth}
\end{figure}

To provide further insight into the multi-NARP scheme, the calculated exciton inversion for the two-notch case is shown in Fig.~\ref{holewidth} for three different notch widths. Three regimes may be identified: (i) for notch spacings less than the notch width, there is effectively a single bin and robust inversion occurs within the single NARP regime \cite{Wilbur:2022}; (ii) for notch spacings similar to the notch width, the bins are not separable and quantum state inversion is incomplete; (iii) for all larger notch spacings, robust quantum state inversion occurs within the multi-NARP regime.  The persistence of adiabatic inversion for a wide range of laser pulse parameters (pulse area, notch width/position, chirp) would facilitate the implementation of the multi-NARP approach for multiplexing in quantum networks. 

\section{Summary and Conclusion}

The multi-NARP triggering method enables simultaneous, high-fidelity inversion of many spectrally distinct QDs, permitting wavelength division multiplexing for high-bandwidth quantum communication.  The robustness of ARP-based schemes for triggering SPSs, including NARP and multi-NARP, is an advantage over resonant laser driving using Rabi rotation \cite{Kuhlmann:2015} and recently proposed quasi-resonant laser pumping schemes using bichromatic pulses \cite{He:2019,Koong:2021,Bracht:2022} for which the inversion fidelity is highly sensitive to the pulse parameters \cite{Koong:2021,Karli:2022}.  Phonon-assisted excitation \cite{Glassl:2013,Reiter:2012} also provides a robust approach to trigger single-photon emission, but in contrast to multi-NARP, robustness comes at the expense of incoherent dynamics that limit the photon indistinguishability. The ability to suppress excitation-induced dephasing tied to electron-phonon coupling in multi-NARP using a positively chirped pulse is an additional advantage over competing approaches \cite{Vanucci:2023}.  The number of channels that could ultimately be encoded depends on the transform-limited pulse duration and the width of the notches.  A comparison of Fig.~\ref{5hole} and Fig.~S3 shows that the number of multiplexed channels can be extended for a given notch width by using wider bandwidth pulses.  The propagation of the wide-bandwidth triggering pulse through an optical fiber can provide a passive approach to introducing the necessary chirp for ARP. If sufficient chirp cannot be imparted in this way, a chirped fiber Bragg grating is an alternative passive approach that can simplify the implementation of the protocol.


In this paper, we have proposed a laser driving scheme that extends the recently developed NARP scheme to the simultaneous triggering of multiple SPSs with a range of emission wavelengths, laying the groundwork for the implementation of wavelength division multiplexing approaches in quantum optical networks.  Our simulations demonstrate the robustness of this triggering scheme to variations in the laser pulse parameters, a feature of all ARP based approaches to quantum state inversion in SPSs, which will facilitate the experimental implementation of multi-NARP.  We also show that excitation-induced dephasing tied to coupling of the exciton with LA phonons may be suppressed through the use of positive chirp in the triggering laser pulse.  Together with the direct excitation of the ground state exciton using the multi-NARP approach, this will enable simultaneous optimization of the key metrics (indistinguishability, single-photon purity, and brightness) for high-performance SPSs. We demonstrated quantum state initialization using the multi-NARP scheme experimentally on two spectrally distinct, telecom-compatible InGaAs QDs.  Our simulations show that femtosecond triggering pulses in conjunction with multi-NARP would provide at least a ten-fold enhancement in the bandwidth of quantum communication systems.   Our work enables the realization of bright, multiplexed QD SPSs and represents an important step towards high-bandwidth quantum communication systems with true single photons, improving upon current approaches using attenuated coherent states.

\begin{backmatter}
\bmsection{Funding}
This research was supported by the Natural Sciences and Engineering Research Council of Canada, Grant No. RGPIN-2020-06322, and the National Research Council of Canada Internet of Things: Quantum Sensors Challenge Programs QSP 042 and QSP 083.


\bmsection{Disclosures}
The authors declare no conflict of interest.

\bmsection{Data Availability Statement}
All data generated or analyzed during this study are included in this published article.


\end{backmatter}


\begin{thebibliography}

\bibitem{Kolodynski:2020} J.~Ko\l{}ody\'{n}ski, A.~M\'{a}ttar, P.~Skrzypczyk, E.~Woodhead, D.~Cavalcanti, K.~Banaszek, and A.~Ac\'{i}n, \enquote{Device-independent quantum key distribution with single-photon sources,} {\protect\JournalTitle{Quantum}} \textbf{4}, 260 (2020).

\bibitem{Lu:2021} C.-Y.~Lu and J.-W. Pan, \enquote{Quantum-dot single photon sources for the quantum internet,} {\protect\JournalTitle{Nat. Nanotechnol.}} \textbf{16}, 1294--1296 (2021). 

\bibitem{Rodt:2021} S.~Rodt and S.~Reitzenstein, \enquote{Integrated nanophotonics for the development of fully functional quantum circuits based on on-demand single-photon emitters,} {\protect\JournalTitle{APL Photonics}} \textbf{6}, 010901 (2021).

\bibitem{Zadeh:2016} I.~E.~Zadeh, A.~W.~Elshaari, K.~D.~J\"{o}ns, \textit{et al.}, \enquote{Deterministic Integration of Single Photon Sources in Silicon Based Photonic Circuits,} {\protect\JournalTitle{Nano Lett.}} \textbf{16}, 2289--2294 (2016).

\bibitem{Bozzio:2022} M.~Bozzio, M.~Vyvlecka, M.~Cosacchi, C.~Nawrath, T.~Seidelmann, J.~C.~Loredo, S.~L.~Portalupi, V.~M.~Axt, P.~Michler, and P.~Walther, \enquote{Enhancing quantum cryptography with quantum dot single-photon sources,} {\protect\JournalTitle{npj Quantum Inf.}} \textbf{8}, 104 (2022).


\bibitem{Chaiwongkhot:2020} P.~Chaiwongkhot, S.~Hosseini, A.~Ahmadi, B.~L.~Higgins, D.~Dalacu, P.~J.~Poole, R.~L.~Williams, M.~E.~Reimer, T.~Jennewein, \enquote{Enhancing secure key rates of satellite QKD using a quantum dot single-photon source,} {\protect\JournalTitle{arXiv}}, 2009.11818 (2020).

\bibitem{Zhang:2025} Y.~Zhang, X.~Ding, Y.~Li, \textit{et al.}, \enquote{Experimental Single-Photon Quantum Key Distribution Surpassing the Fundamental Weak Coherent-State Rate Limit,} {\protect\JournalTitle{Phys. Rev. Lett.}} \textbf{134}, 210801 (2025).


\bibitem{Deiteil:2017} A.~Delteil, Z.~Sun, S.~Fält, and A.~Imamo\u{g}lu, \enquote{Realization of a Cascaded Quantum System: Heralded Absorption of a Single Photon Qubit by a Single-Electron Charged Quantum Dot,} {\protect\JournalTitle{Phys. Rev. Lett.}} \textbf{118}, 177401 (2017).


\bibitem{Anderson:2020} M.~Anderson, T.~M\"{u}ller, J.~Skiba-Szymanska1, A.~B.~Krysa, J.~Huwer, R.~M.~Stevenson, J.~Heffernan, D.~A.~Ritchie, and A.~J.~Shields, \enquote{Gigahertz-Clocked Teleportation of Time-Bin Qubits with a Quantum Dot in the Telecommunication C Band,} {\protect\JournalTitle{Phys. Rev. Applied}} \textbf{13}, 054052 (2020).

\bibitem{Maring:2024} N.~Maring, A.~Fyrillas, M.~Pont \textit{et al.}, \enquote{A versatile single-photon-based quantum computing platform,} {\protect\JournalTitle{Nat. Photonics}} \textbf{18}, 603--609 (2024).

\bibitem{Thoma:2024} S.~E.~Thomas, L.~Wagner, R.~Joos, \textit{et al.},  \enquote{Deterministic storage and retrieval of telecom light from a quantum dot single-photon source interfaced with an atomic quantum memory,} {\protect\JournalTitle{Sci. Adv.}} \textbf{10}, eadi7346 (2024).


\bibitem{Wells:2023} L.~Wells, T.~M\"{ü}ller, R.~M.~Stevenson, J.~Skiba-Szymanska, D.~A.~Ritchie, and A.~J.~Shields, \enquote{Coherent light scattering from a telecom C-band quantum dot,} {\protect\JournalTitle{Nat. Commun.}} \textbf{14}, 8371 (2023).

\bibitem{Senellart:2017} P.~Senellart, G.~Solomon and A.~White, \enquote{High-performance semiconductor quantum-dot single-photon sources,}  {\protect\JournalTitle{Nat. Nanotechnol.}} \textbf{12}, 1026--1039 (2017).


\bibitem{Somaschi:2016} N.~Somaschi, V.~Giesz, L.~De~Santis, \textit{et al.},  \enquote{Near-optimal single-photon sources in the solid state,} {\protect\JournalTitle{Nature Photon.}} \textbf{10}, 340 (2016).

\bibitem{Schweickert:2018} L.~Schweickert, K.~D.~J\"{o}ns, K.~D.~Zeuner, S.~F.~Covre~da~Silva; H.~Huang, T.~Lettner, M.~Reindl, J.~Zichi; R.~Trotta; A.~Rastelli, V.~Zwiller, \enquote{On-demand generation of background-free single photons from a solid-state source,} {\protect\JournalTitle{Appl. Phys. Lett.}} \textbf{112}, 093106 (2018).

\bibitem{Karli:2024} Y.~Karli, D.~A.~Vajner, F,~Kappe, \textit{et al.}, \enquote{Controlling the photon number coherence of solid-state quantum light sources for quantum cryptography,}  {\protect\JournalTitle{npj Quantum Inf.}} \textbf{10}, 17 (2024).



\bibitem{Haffouz:2018} S.~Haffouz, K.~D.~Zeuner, D.~Dalacu, \textit{et al.}, \enquote{Bright Single InAsP Quantum Dots at Telecom Wavelengths in Position-Controlled InP Nanowires: The Role of the Photonic Waveguide,} {\protect\JournalTitle{Nano Lett.}},  \textbf{18}, 3047--3052 (2018).

\bibitem{Vajner:2024} D.~A.~Vajner, P.~Holewa, E.~Zi\c{e}ba-Ost\'{o}j, \textit{et al.}, \enquote{On-Demand Generation of Indistinguishable Photons in the Telecom C-Band Using Quantum Dot Devices,} {\protect\JournalTitle{ACS Photonics}},  \textbf{11}, 339--347 (2024).

\bibitem{Kim:2025} J.~Kim, J.~Kaupp, Y.~Reum, \textit{et al.}, \enquote{Two-Photon Interference from an InAs Quantum Dot emitting in the Telecom C-Band, }  {\protect\JournalTitle{arXiv}}, 2501.15970 (2025).

\bibitem{Nawrath:2023} C.~Nawrath, R.~Joos, S.~Kolatschek, \textit{et al.}, \enquote{Bright Source of Purcell-Enhanced, Triggered, Single Photons in the Telecom C-Band,} {\protect\JournalTitle{Adv. Quantum Technol.}},  \textbf{6}, 2300111 (2023).

\bibitem{Anderson:2021} M.~Anderson, T.~M\"{u}ller, J.~Skiba-Szymanska, \textit{et al.},  \enquote{Coherence in single photon emission from droplet epitaxy and Stranski–Krastanov quantum dots in the telecom C-band,} {\protect\JournalTitle{Appl. Phys. Lett.}},  \textbf{118}, 014003 (2021).


\bibitem{Glassl:2013} M.~Gl\"{a}ssl, A.~M.~Barth, and V.~M.~Axt, \enquote{Proposed Robust and High-Fidelity Preparation of Excitons and Biexcitons in Semiconductor Quantum Dots Making Active Use of Phonons,} {\protect\JournalTitle{Phys. Rev. Lett.}},  \textbf{110}, 147401 (2013).

\bibitem{Reiter:2012} D.~E.~Reiter, S.~L\"{u}ker, K.~Gawarecki, \textit{et al.}, \enquote{Phonon Effects on Population Inversion in Quantum Dots: Resonant, Detuned and Frequency-Swept Excitations,} {\protect\JournalTitle{Acta Physica Polonica A}},  \textbf{122}, 1065--1068 (2012).

\bibitem{Huber:2020} T.~Huber, M.~Davanco, M.~M\"{u}ller, \textit{et al.}, \enquote{Filter-free single-photon quantum dot resonance fluorescence in an integrated cavity-waveguide device,} {\protect\JournalTitle{Optica}},  \textbf{7}, 380--385 (2020).

\bibitem{Uppu:2020} R.~Uppu, H.~T.~Eriksen, H.~Thyrrestrup, \textit{et al.}, \enquote{On-chip deterministic operation of quantum dots in dual-mode waveguides for a plug-and-play single-photon source,} {\protect\JournalTitle{Nat. Commun.}},  \textbf{11}, 3782 (2020).  
\bibitem{Santori:2001} C.~Santori, M.~Pelton, G.~Solomon, Y.~Dale, and Y.~Yamamoto, \enquote{Triggered Single Photons from a Quantum Dot,} {\protect\JournalTitle{Phys. Rev. Lett.}}, \textbf{86}, 1502 (2001).

\bibitem{Mathew:2014} R.~Mathew, E.~Dilcher, A.~Gamouras, \textit{et al.}, \enquote{Subpicosecond adiabatic rapid passage on a single semiconductor quantum dot: Phonon-mediated dephasing in the strong-driving regime,}  {\protect\JournalTitle{Phys. Rev. B}}, \textbf{90}, 035316 (2014).

\bibitem{Wei:2014} Y.-J.~Wei, Y.-M.~He, M.-C.~Chen, \textit{et al.}, \enquote{Deterministic and Robust Generation of Single Photons from a Single Quantum Dot with 99.5\% Indistinguishability Using Adiabatic Rapid Passage,} {\protect\JournalTitle{Nano Lett.}}, \textbf{14}, 6515--6519 (2014).

\bibitem{He:2019} Y.-M.~He, H.~Wang, C.~Wang, \textit{et al.}, \enquote{Coherently driving a single quantum two-level system with dichromatic laser pulses,} {\protect\JournalTitle{Nat. Phys.}}, \textbf{15}, 941--946 (2019).

\bibitem{Koong:2021} Z.~X.~Koong, E.~Scerri, M.~Rambach \textit{et al.}, \enquote{Coherent Dynamics in Quantum Emitters under Dichromatic Excitation,} {\protect\JournalTitle{Phys. Rev. Lett.}}, \textbf{126}, 047403 (2021).

\bibitem{Bracht:2022} T.~K.~Bracht, M.~Cosacchi, T.~Seidelmann, \textit{et al.}, \enquote{Swing-Up of Quantum Emitter Population Using Detuned Pulses,} {\protect\JournalTitle{PRX Quantum}}, \textbf{2}, 040354 (2021).

\bibitem{Wilbur:2022} G.~R.~Wilbur, A.~Binai-Motlagh, A.~Clarke, \textit{et al.}, \enquote{Notch-filtered adiabatic rapid passage for optically driven quantum light sources,}  {\protect\JournalTitle{APL Photonics}}, \textbf{7}, 111302 (2022).

\bibitem{Kuhlmann:2013} A.~V.~Kuhlmann, J.~Houel, D.~Brunner, A.~Ludwig, D.~Reuter, A.~D.~Wieck, and R.~J~Warburton, \enquote{A dark-field microscope for background-free detection of resonance fluorescence from single semiconductor quantum dots operating in a set-and-forget mode,} {\protect\JournalTitle{Rev. Sci. Instrum.}}, \textbf{84},073905 (2013).

\bibitem{Kaldewey:2017} T.~Kaldewey, S.~L\"{u}ker, A.~V.~Kuhlmann, \textit{et al.}, \enquote{Demonstrating the decoupling regime of the electron-phonon interaction in a quantum dot using chirped optical excitation,} {\protect\JournalTitle{Phys. Rev. B}}, \textbf{95}, 241306(R) (2017).

\bibitem{Ramachandran:2020} A.~Ramachandran, G.~R.~Wilbur, S.~O’Neal, D.~G.~Deppe, and K.~C.~Hall, \enquote{Suppression of decoherence tied to electron–phonon coupling in telecom-compatible quantum dots: low-threshold reappearance regime for quantum state inversion,}  {\protect\JournalTitle{Optics Lett.}}, \textbf{45}, 6498--6501 (2020).
 
\bibitem{Beraza:2025} I.~Beraza, M.~Zahidy, R.~Mueller, \textit{et al.}, \enquote{Quantum communication multiplexing in LP-modes enabled by photonic lanterns,} {\protect\JournalTitle{arXiv}}, 2502.12865 (2025).

\bibitem{Lio:2018} B.~Da~Lio, D.~Bacco, D.~Cozzolino, \textit{et al.}, \enquote{Record-High Secret Key Rate for Joint Classical and Quantum Transmission Over a 37-Core Fiber,} {\protect\JournalTitle{2018 IEEE Photonics Conference (IPC)}}, Reston, VA, USA, 1--2,(2018). 

\bibitem{Koong:2020} Z.-X. Koong, G.~Ballesteros-Garcia, R.~Proux, \textit{et al.}, \enquote{Multiplexed Single Photons from Deterministically Positioned Nanowire Quantum Dots,}  {\protect\JournalTitle{Phys. Rev. Appl.}}, \textbf{14}, 034011 (2020).

\bibitem{Eriksson:2019} T.~A.~Eriksson, T.~Hirano, B.~J.~Puttnam, \textit{et al.}, \enquote{Wavelength division multiplexing of continuous variable quantum key distribution and 18.3 Tbit/s data channels,}   {\protect\JournalTitle{Commun. Phys.}}, \textbf{2}, 9 (2019).

\bibitem{Zahidy:2022} M.~Zahidy, Y.~Liu, D.~Cozzolino, \textit{et al.}, \enquote{Photonic integrated chip enabling orbital angular momentum multiplexing for quantum communication,} {\protect\JournalTitle{Nanophotonics}}, \textbf{11}, 821--827 (2022).

\bibitem{Park:2022} C.~H.~Park, M.~K.~Woo, B.~K.~Park, \textit{et al.}, \enquote{2×N twin-field quantum key distribution network configuration based on polarization, wavelength, and time division multiplexing,}  {\protect\JournalTitle{npj Quantum Inf.}}, \textbf{8}, 48 (2022).

\bibitem{Ramachandran:2021} A.~Ramachandran, J.~Fraser-Leach, S.~O'Neal, D.~G.~Deppe, and K.~C.~Hall, \enquote{Experimental quantification of the robustness of adiabatic rapid passage for quantum state inversion in semiconductor quantum dots,}  {\protect\JournalTitle{Optics Express}}, \textbf{29}, 41766--41775 (2021).

\bibitem{Ramachandran:2024} A.~Ramachandran, G.~R.~Wilbur, R.~Mathew, \textit{et al.}, \enquote{Robust parallel laser driving of quantum dots for multiplexing of quantum light sources,} {\protect\JournalTitle{Sci. Rep.}}, \textbf{14}, 5356 (2024).


\bibitem{Remesh:2023} V.~Remesh, R.~G.~Kr\"{a}mer, R. Schwarz, \textit{et al.}, \enquote{Compact chirped fiber bragg gratings for single-photon generation from quantum dots,} {\protect\JournalTitle{APL Photonics}}, \textbf{8}, 101301 (2023).

\bibitem{Ramsay:2011} A.~J.~Ramsay, T.~M.~Godden, S.~J.~Boyle, \textit{et al.}, \enquote{Effect of detuning on the phonon induced dephasing of optically driven InGaAs/GaAs quantum dots,} {\protect\JournalTitle{J. Appl. Phys.}}, \textbf{109}, 102415 (2011).

\bibitem{Kuhn:2012} S.~L\"{u}ker, K.~Gawarecki, D.~E.~Reiter, A.~Grodecka-Grad, V.~M.~Axt, P.~Machnikowski, and T.~Kuhn, \enquote{Influence of acoustic phonons on the optical control of quantum dots driven by adiabatic rapid passage,} {\protect\JournalTitle{Phys. Rev B}}, \textbf{85}, 121302(R) (2012).

\bibitem{Debnath:2012} A.~Debnath, C.~Meier, B.~Chatel, and T.~Amand, \enquote{Chirped laser excitation of quantum dot excitons coupled to a phonon bath,} {\protect\JournalTitle{Phys. Rev B}}, \textbf{86}, 161304(R) (2012).

\bibitem{Shore:book} B.~W.~Shore, \enquote{Manipulating Quantum Structures Using Laser Pulses,}  (Cambridge University Press, 2011).

\bibitem{Gamouras:2013} A.~Gamouras, R.~Mathew, S.~Freisem, D.~G.~Deppe, K.~C.~Hall, \enquote{Simultaneous Deterministic Control of Distant Qubits in Two Semiconductor Quantum Dots,} {\protect\JournalTitle{Nano Lett.}}, \textbf{13}, 4666 (2013).

\bibitem{Kuhlmann:2015} A.~V.~Kuhlmann, J.~H.~Prechtel, J.~Houel, A.~Ludwig, D.~Reuter, A.~D.~Wieck, and R. J. Warburton, \enquote{Transform-limited single photons from a single quantum dot,} {\protect\JournalTitle{Nat. Commun.}}, \textbf{6}, 8204 (2015).

\bibitem{Karli:2022} Y.~Karli, F.~Kappe, V.~Remesh, \textit{et al.} \enquote{SUPER Scheme in Action: Experimental Demonstration of Red-Detuned Excitation of a Quantum Emitter,} {\protect\JournalTitle{Nano Lett.}}, \textbf{22}, 6567--6572 (2022).

\bibitem{Vanucci:2023} L.~Vannucci, and N.~Gregersen, \enquote{Highly efficient and indistinguishable single-photon sources via phonon-decoupled two-color excitation,} {\protect\JournalTitle{Phys. Rev. B}}, \textbf{107}, 195306 (2023).

\bibitem{Htoon:2002} H.~Htoon, T.~Takagahara, D.~Kulik, \textit{et al.}, \enquote{Interplay of Rabi oscillations and quantum interference in semiconductor quantum dots,}  {\protect\JournalTitle{Phys. Rev. Lett.}}, \textbf{88}, 087401 (2002).






 \end{thebibliography}
\end{document}